\documentclass[11pt,epsf]{article}
\usepackage{amsmath}
\usepackage{amsfonts}
\usepackage{amssymb}
\usepackage{graphicx}
\usepackage{color}
\usepackage{comment}
\newcommand {\dfn} {\stackrel{\Delta} {=}}

\newcommand {\reals} {{\rm I\!R}}

\newcommand {\bE} {\mbox{\boldmath $E$}}

\newcommand{\calN}{{\cal N}}

\newcommand{\calU}{{\cal U}}

\begin{document}
\thispagestyle{empty}
\title{Some Families of Jensen-like Inequalities\\ 
with Application to Information Theory
%\thanks{This research was supported by my wife and kids.}
}
\author{Neri Merhav
%\thanks{
%Currently on sabbatical leave at HP Laboratories,
%1501 Page Mill Road, MS 3U-4, Palo Alto CA 94304, USA.}
}
\date{}
\maketitle

\begin{center}
The Andrew \& Erna Viterbi Faculty of Electrical and Computer Engineering\\
Technion - Israel Institute of Technology \\
Technion City, Haifa 32000, ISRAEL \\
E--mail: {\tt merhav@technion.ac.il}\\
\end{center}
\vspace{1.5\baselineskip}
\setlength{\baselineskip}{1.5\baselineskip}

\begin{center}
In memory of Jacob Ziv,\\
a shining star in the sky of information theory\\
and a great inspiration to many of us, for years to come.
\end{center}

\vspace{0.5cm}

\begin{abstract}
It is well known that the traditional Jensen inequality is proved by lower bounding the
given convex function, $f(x)$, by the tangential affine function that passes
through the point $(\bE\{X\},f(\bE\{X\}))$, where $\bE\{X\}$ is the
expectation of the random variable $X$. While this tangential affine
function yields the tightest
lower bound among all lower bounds induced by affine functions that are
tangential to $f$, it turns out that when the function $f$ is just part of a more
complicated expression whose expectation is to be bounded, the tightest lower
bound might belong to a tangential
affine function that passes through a point different than
$(\bE\{X\},f(\bE\{X\}))$. In this paper, we take advantage of this
observation, by optimizing the point of tangency with regard to the specific
given expression, in a variety of cases, and thereby derive several families of inequalities, henceforth referred to as
``Jensen-like'' inequalities, which are new to the best knowledge of the
author. The degree of tightness and the potential usefulness of these inequalities is demonstrated in
several application examples related to information theory.
\end{abstract}

\section{Introduction}

As is well known, the Jensen inequality is one of the most fundamental
and useful mathematical tools in a variety of fields,
including information theory. Interestingly, it includes
many other very well known inequalities, which are
important on their own, as special cases. Among many examples, we mention
the Shwartz-Cauchy
inequality (which in turn supports uncertainty principles and the
Cram\'er--Rao bound), the Lyapunov inequality, the H\"older inequality, 
and the inequalities among the harmonic, geometric and arithmetic
means. In the field of 
information theory, the Jensen inequality stands at the basis of the information inequality
(i.e., the non-negativity of
the relative entropy), the data processing inequality (which in
turn, leads to the Fano inequality), and the inequality between conditional
and unconditional entropies.
Moreover, it plays a central role in support of
the derivation of single--letter formulas in Shannon theory and in the
theory of maximum entropy under moment constraints (see, for example, Chapter
12 of \cite{CT06}).

During the last two decades, there have been many research efforts around
Jensen's inequality, which included refinements \cite{XL20}, \cite{DUKIW21},
\cite{WKSS22}, \cite{SBS23}, variations
\cite{JAHM20}, \cite{MP07}, \cite{BMP08}, improvements \cite{SG12}, \cite{Walker14},
\cite{LB18}, and extensions \cite{SA21},
just to name a few. There have also many derivations of
reversed versions of the Jensen inequality, see, e.g.,
\cite{JP00}, \cite{BDP01}, \cite{Simic09a}, \cite{Simic09b}, \cite{Dragomir10},
\cite{Dragomir13}, \cite{ KKC20a}, \cite{KKC20b}, \cite{WGFS21}, \cite{ABZ21},
\cite{BAT20} and \cite{me22},
for a non--exhaustive sample of articles.
In most of them, the derived inequalities are
exemplified in many applications, for instance, useful relationships
between arithmetic and geometric means, converse bounds on the entropy, the
relative entropy, as well as the more general $f$-divergence, converse forms
of the H\"{o}lder inequality, and so on. In many of these works,
the main results are given in the form of an upper bound on the
difference, $\bE\{f(X)\}-f(\bE\{X\})$, where $f$ is a convex function,
$\bE\{\cdot\}$ is
the expectation operator, and $X$ is the random variable. But those bounds,
depend mostly on global parameters associated with $f$, for example, its
range and domain, but not particularly on the underlying probability function
(probability density function in the continuous case, or 
probability mass function in the discrete case), 
of $X$. For one thing, a desirable property of a reverse Jensen inequality
would be that it is tight when
$X$ is well concentrated in the vicinity of its mean, just like
the same well known property of the ordinary Jensen inequality.
In \cite{me22}, there is an attempt to address this issue.

This paper revisits the Jensen inequality from a different angel.
It is based on the following simple observation, which is rooted in the proof
of Jensen's inequality: The given convex function, $f(x)$, is lower bounded by
the tangential affine function, $\ell(x)=f(a)+f'(a)(x-a)$, where $a$ is an
arbitrary number in the domain of $x$ and $f'(a)$ is the derivative of $f$
at $x=a$ (provided that $f$ is differentiable at $x=a$). By
selecting $a=\bE\{X\}$ and taking expectations of both sides of the inequality,
$f(X)\ge\ell(X)$,
the Jensen inequality is readily proved. The point to be remembered
is that here, $a_*=\bE\{X\}$ is the {\em optimal choice of} $a$ in the sense of maximizing
$\bE\{\ell(X)\}$ over all possible values of $a$, thus yielding the tightest lower bound within this class of
lower bounds on $\bE\{f(X)\}$. The optimal choice of $a$, however, might be different than
$\bE\{X\}$ when the function $f(X)$ is only a part of a more complicated
expression whose expectation is to be lower bounded. For example, one might be
interested in lower bounding $\bE\{g[f(X)]\}$, where $g$ is a monotonically
non-decreasing function, or $\bE\{f(X)g(X)\}$, where $g$ is  positive and/or convex
function, or a combination of both, etc. 

To demonstrate this fact, consider the
example (to be treated in detail in Section \ref{s1}) 
of lower bounding $\bE\{f(X)g(X)\}$, where $g$ is a positive function.
In this case,
\begin{equation}
\bE\{f(X)g(X)\}\ge\bE\{[f(a)+f'(a)(X-a)]g(X)\},
\end{equation}
and by maximizing the right--hand side (r.h.s.) over $a$, we
easily obtain that the optimal choice of $a$ here is
$a_*=\bE\{Xg(X)\}/\bE\{g(X)\}$, yielding the inequality,
\begin{equation}
\bE\{f(X)g(X)\}\ge
f\left(\frac{\bE\{Xg(X)\}}{\bE\{g(X)\}}\right)\cdot\bE\{g(X)\},
\end{equation}
which is useful as long as $g$ is such that we can easily calculate both $\bE\{g(X)\}$
and $\bE\{Xg(X)\}$. While this particular inequality could have been obtained also by
applying the (ordinary) Jensen inequality, $\bE\{f(X)\}\ge f(\bE\{X\})$,
with respect to (w.r.t.) the density,
$\tilde{p}(x)=p(x)g(x)/\int_{-\infty}^\infty p(x')g(x')\mbox{d}x'$, we will
see in the sequel, also various examples of inequalities with no apparent simple
interpretations such as this. We henceforth refer to these classes of inequalities as {\em
Jensen-like inequalities} since they are derived using the same general idea
that underlies the proof the classical Jensen inequality.
We will also demonstrate the usefulness of these
inequalities in information theory. 

Our contributions, in this work, have the following
features:
\begin{enumerate}
\item In many cases (like the one above), the optimal value of the parameter(s)
(e.g., the parameter $a$ in the above discussion) can be found
in closed form. In other cases, the resulting expressions may not lend
themselves to closed-form optimization, and then we have two possibilities:
(i) carry out the optimization numerically, and (ii) select an arbitrary
choice of $a$ and obtain a valid lower bound, bearing in mind that an educated
guess can potentially result in a good bound.
\item Our inequalities provide two types of bounds: (i)  bounds that require the calculation
of the first two moments (or equivalently, the first two cumulants) of $X$, and (ii) bounds that require
the calculation of the moment generating function (MGF) of $X$ and its
derivative, or equivalently, the cumulant generating function (CGF) of $X$ and
its derivative. All these types of moments are often easily calculable in
closed form,
especially, in situations where $X$ is given by the sum of independent and identically
distributed (i.i.d.) random variables, which is frequently encountered in
information-theoretic applications.
\item Most of our derivations extend to convex functions of more than one
variable.
\item The classes of Jensen-like inequalities that we consider allow enough
flexibility to obtain derivations of lower bounds on functions that are not
necessarily convex, and even for some concave functions, and thereby open the
door for another route to reverse Jensen inequalities. This can be
accomplished by representing the given function in one of the categories
discussed (e.g., a product of a convex function and a positive function, a product
of two positive convex functions, a composition of a monotone function and a
convex function, etc.).
\item We demonstrate the utility of the Jensen-like inequalities in several examples of
information-theoretic relevance. We also display numerical results that
exemplify the degree
of tightness of these bounds.
\item Our Jensen-like inequalities have the desirable property of becoming tighter as $X$
becomes more and more concentrated around
its mean, just like the ordinary Jensen inequality.
\item Throughout the paper, we confine ourselves to lower bounds on expectations of
expressions that include a convex function $f$, but it should be understood
that they all continue to apply also if $f$ is concave and the inequalities are
reversed.
\item It should be understood that the classes of Jensen-like
inequalities that we derive in this work are just examples that demonstrate the basic
underlying idea of optimizing the point of tangency to the given convex 
function for the specific expression at hand. It is conceivable that the same idea can be applied to many more
situations of theoretical and practical interest.
\end{enumerate}

Finally, a technical comment is in order: In all forthcoming derivations, it
will be assumed
that the convex function $f$ is differentiable at least at the optimal point,
$x=a_*$. It should be understood, however, that this assumption is made mainly for convenience, not really
because of necessity. As is well known, convex functions always have at least
one-sided derivatives at any point, and then $f'(a_*)$ can be taken to be any
value between the left derivative and the right derivative of $f$ at $x=a_*$. 
In order to show that the point of zero-derivative of the lower bound (w.r.t.\
$a$) indeed yields a maximum (and not a minimum, etc.) of
the lower bound, we will need to further assume that $f$ is twice differentiable, but
such an assumption will not limit the applicability of the claimed lower
bound, because the lower bound applies to any value of $a$, including the
point of zero-derivative, even if this point cannot be proved to yield the maximum of the
lower bound using the standard methods. Similar comments apply when the lower bound will depend on more
than one parameter.

In the remaining part of this article, each section is devoted to a different
class of Jensen-like inequalities, which corresponds to a different form of an
expression that includes the convex function, $f$.

\section{A product of a convex function and a positive function}
\label{s1}

In this section, we focus on lower bounding expressions of the form
$\bE\{f(X)g(X)\}$, where $f$ is convex and $g$ is non-negative.
Indeed, let $f:\reals\to\reals$ be a  
convex function and let $g:\reals\to\reals^+$ be a non-negative function.
Then, for any $a\in\reals$,
\begin{eqnarray}
	\bE\{f(X)g(X)\}&\ge&\bE\{[f(a)+f'(a)(X-a)]g(X)\}\\
	&=&[f(a)-af'(a)]\bE\{g(X)\}+f'(a)\bE\{Xg(X)\}.
\end{eqnarray}
To find the value of $a$ that maximizes the r.h.s., we equate the derivative to zero, and get:
\begin{equation}
	[f'(a)-f'(a)-af''(a)]\bE\{g(X)\}+f''(a)\bE\{Xg(X)\}=0
\end{equation}
or equivalently,
\begin{equation}
	f''(a)[\bE\{Xg(X)\}-a\bE\{g(X)\}]=0,
\end{equation}
which yields 
\begin{equation}
	a_*=\frac{\bE\{Xg(X)\}}{\bE\{g(X)\}},
\end{equation}
and it is easy to verify that the second derivative at $a=a_*$ 
is $-f''(a_*)\bE\{g(X)\}<0$, which means that it is a maximum (at least a
local one).
The resulting lower bound on $\bE\{f(X)g(X)\}$ is then given by
\begin{equation}
\label{ineq1}
	\bE\{f(X)g(X)\}\ge f\left(\frac{\bE\{Xg(X)\}}{\bE\{g(X)\}}\right)\cdot\bE\{g(X)\}.
\end{equation}
This result extends straightforwardly to the case where $X$ is a vector,
provided that $f$ is jointly convex in all components of $X$. In particular,
it extends to the case where $f$ and $g$ act of different random variables,
$X$ and $Y$, with a joint distribution:
\begin{equation}
	\bE\{f(X)g(Y)\}\ge f\left(\frac{\bE\{Xg(Y)\}}{\bE\{g(Y)\}}\right)\cdot\bE\{g(Y)\}.
\end{equation}

We next consider several examples.\\

\noindent
{\em Example 1.}
Let $f(x)=-\ln x$ and $g(x)=x$, $x> 0$. Applying inequality
(\ref{ineq1}),
\begin{equation}
\bE\{-X\ln X\}\ge
-\bE\{X\}\cdot\ln\frac{\bE\{X^2\}}{\bE\{X\}}=-\bE\{X\}\cdot\ln(\bE\{X\})-
\bE\{X\}\cdot\ln\left(1+\frac{\mbox{Var}\{X\}}{[\bE\{X\}]^2}\right).
\end{equation}
Note that the function $-x\ln x$ is concave, rather than convex, yet we have
here a lower bound
(rather than an upper bound) to its expectation, namely, a reversed Jensen
inequality. The first term on the right-most side is the (ordinary) Jensen upper bound 
on $\bE\{-X\ln X\}$,
and the second term is the gap, which
depends not only on the expectation of $X$, but also on its variance, which manifests the
fluctuations around $\bE\{X\}$.
This inequality has an immediate
application for obtaining a lower bound to the expectation of the empirical entropy of a sequence
drawn by a memoryless source, which is relevant in the context of universal
source coding \cite{KT81}.
Each term of the empirical entropy is of the form $-X\ln X$, where
$X=N(u)/N$, $N(u)$ being the number of occurrences of a letter $u$ in a randomly drawn $N$-tuple from a
a memoryless source, $P$, with a finite alphabet, $\calU$. Clearly, each $N(u)$ is a binomial random variable
with $N$ trials and probability of success, $P(u)$.
In this case, $\bE\{X\}=P(u)$ and 
$\mbox{Var}\{X\}=P(u)[1-P(u)]/N$.
Thus, denoting the entropy and the empirical entropy, respectively, by
\begin{eqnarray}
H&=&-\sum_{u\in\calU}P(u)\ln P(u)\\
\hat{H}&=&-\sum_{u\in\calU}\frac{N(u)}{N}\ln\left(\frac{N(u)}{N}\right),
\end{eqnarray}
with the convention that $0\ln 0\dfn 0$, we have:
\begin{eqnarray}
\bE\{\hat{H}\}&\ge&-\sum_{u\in\calU}P(u)\ln
P(u)-\sum_{u\in\calU}P(u)\ln\left[1+\frac{P(u)[1-P(u)]/N}{P^2(u)}\right]\nonumber\\
&=& H-\sum_{u\in\calU}P(u)\ln\left(1+\frac{1-P(u)}{NP(u)}\right)\nonumber\\
&\ge& H-\sum_{u\in\calU}P(u)\cdot\frac{1-P(u)}{NP(u)}\nonumber\\
&=& H-\frac{1}{N}\sum_{u\in\calU}[1-P(u)]\nonumber\\
&=&H-\frac{|\calU|-1}{N},
\end{eqnarray}
where $|\calU|$ is the cardinality of $\calU$.
The use of the ordinary Jensen inequality yields an upper bound rather than a lower bound, 
$\bE\{\hat{H}\}\le H$. This completes Example 1. 

\noindent
{\em Example 2.} Let $s$ and $t$ be two real numbers whose difference,
$s-t$, is either negative or larger than unity. Now, let
$g(x)=x^t$, and $f(x)=x^{s-t}$. Then,
\begin{eqnarray}
\bE\{X^s\}&=&\bE\{X^tX^{s-t}\}\nonumber\\
&\ge&\left(\frac{\bE\{X^{t+1}\}}{\bE\{X^t\}}\right)^{s-t}\cdot\bE\{X^t\}\nonumber\\
	&=&\frac{(\bE\{X^{t+1}\})^{s-t}}{(\bE\{X^t\})^{s-t-1}}.
\end{eqnarray}
In particular, for $t=1$ and $s\notin(1,2)$, this becomes
\begin{equation}
	\bE\{X^s\}\ge
\frac{(\bE\{X^2\})^{s-1}}{(\bE\{X\})^{s-2}}=[\bE\{X\}]^s\cdot\left(1+\frac{\mbox{Var}\{X\}}{[\bE\{X\}]^2}\right)^{s-1}
\end{equation}
which is, once again, a bound that depends only on the first two moments of $X$.  
For $s\in(0,1)$, the function $x^s$ is concave, and so, this is a reversed 
version of Jensen inequality. For $s\le 0$ and $s\ge 2$, the function $x^s$ is
convex, and so, this is an improved version of Jensen's inequality: While the
first factor, $[\bE\{X\}]^s$, corresponds to the
ordinary Jensen inequality, the second factor expresses the improvement, which
depends on the relative fluctuation term, $\mbox{Var}\{X\}/[\bE\{X\}]^2$.

To particularize this example even further, consider the problem of randomized guessing under a
distribution $Q$ (see, e.g., \cite{MC20} and many references therein). 
Then, the probability of a single success in guessing a discrete alphabet
random variable, $X$, given that we know (but not the guesser) that $X=x$,
is $Q(x)$. In sequential guessing until the first success, the number of guesses, $G$, is a geometric RV with
parameter $p=Q(x)$, whose mean and variance are $1/p$ and $(1-p)/p^2$,
respectively. For $s\in(1,2)$,
\begin{equation}
\bE\{G^s\}\ge
\left(\frac{1}{p}\right)^s\cdot\left(1+\frac{(1-p)/p^2}{1/p^2}\right)^{s-1}=
\frac{(2-p)^{s-1}}{p^s}=\frac{[2-Q(x)]^{s-1}}{[Q(x)]^s}.
\end{equation}
This completes Example 2. 

\noindent
{\em Example 3.}
Let $f$ be an arbitrary convex function
and let $g(x)=e^{sx}$, where $s$ is a given real number.
Then, inequality (\ref{ineq1}) becomes:
\begin{equation}
\bE\{f(X)e^{sX}\}\ge f(\psi'(s))\cdot e^{\psi(s)}
\end{equation}
where 
\begin{equation}
\psi(s)=\ln\bE\{e^{sX}\}
\end{equation}
is the CGF of $X$ and $\psi'(s)$ is its derivative.
This gives a lower bound in terms of the CGF of $X$ and its derivative.
The ordinary Jensen inequality is obtained as the special case of $s=0$, where $\psi(0)=0$ and
$\psi'(0)=\bE\{X\}$.

\section{A composition of a monotone function and a convex function}
\label{s2}

Another family of Jensen-like inequalities corresponds to the need to lower
bound an expression of the form
$\bE\{g[f(X)]\}$, where $f$ is convex as before and $g$ is a monotonically non-decreasing function.
The general idea is to carry out the optimization of the r.h.s.\ of the
following inequality.
\begin{equation}
	\bE\{g[f(X)]\}\ge\sup_a\bE\{g[f(a)+f'(a)(X-a)]\}.
\end{equation}
In the important special case where $g(x)=e^{x}$, we have:
\begin{eqnarray}
\bE\{e^{f(X)}\}&\ge&\sup_a\bE\{e^{f(a)+f'(a)(X-a)}\}\nonumber\\
&=&\sup_ae^{f(a)-af'(a)}\bE\{e^{Xf'(a)}\}\nonumber\\
	&=&\exp\left\{\sup_a\{f(a)-af'(a)+\psi[f'(a)]\}\right\},
\end{eqnarray}
where $\psi(\cdot)$ is again the CGF of $X$. 
The optimal value, $a_*$, of $a$, is the solution to the equation
obtained by equating the derivative of the exponent to zero, i.e., 
\begin{equation}
	\psi'[f'(a_*)]=a_*,~~\mbox{provided that}~~f''(a_*)\psi''[f'(a_*)]<1,
\end{equation}
where $\psi'(\cdot)$ and $\psi''(\cdot)$ are the first and the second
derivatives of $\psi(\cdot)$, respectively.\\

\noindent
{\em Example 4.} Consider the case where
$f(x)=sx^2/2$ and $X\sim\calN(\mu,\sigma^2)$, where $\sigma^2 < 1/s$, as otherwise,
$\bE\{e^{sX^2}\}=\infty$. In this case, the condition
$f''(a_*)\psi''[f'(a_*)]<1$ is equivalent to $\sigma^2 < 1/s$, and we have
$f'(a)=sa$, $\psi(t)=\mu t+\sigma^2t^2/2$, and so, $\psi'(t)=\mu+\sigma^2t$, which means that
$\psi'[f'(a)]=\mu+\sigma^2sa$. 
The equation for the optimal $a$ becomes then
\begin{equation}
	\mu+\sigma^2sa=a,
\end{equation}
whose solution is
\begin{equation}
	a=a_*\dfn\frac{\mu}{1-\sigma^2s},
\end{equation}
which yields
\begin{equation}
\label{jli}
	\bE\left\{e^{sX^2/2}\right\}\ge \exp\left\{sa_*^2/2-sa_*^2+\mu sa_*+\sigma^2s^2a_*^2/2\right\}=
	\exp\left\{\frac{\mu^2s}{2(1-\sigma^2s)}\right\}.
\end{equation}
The ordinary Jensen inequality yields
\begin{equation}
	\bE\left\{e^{sX^2/2}\right\}\ge
\exp\left\{s\bE\{X^2\}/2\right\}=e^{s(\mu^2+\sigma^2)/2},
\end{equation}
which does not capture the singularity at $s=1/\sigma^2$. 
The exact calculation yields

\begin{equation}
	\bE\left\{e^{sX^2/2}\right\}=\frac{1}{\sqrt{1-\sigma^2s}}\cdot
	\exp\left\{\frac{\mu^2s}{2(1-\sigma^2s)}\right\},
\end{equation}
namely, the Jensen-like bound (\ref{jli}) gives the correct exponential term (along with
the singularity at $s=1/\sigma^2$) and differs
from the exact quantity only in the pre-exponential factor.

\section{A product of a convex function and a monotone-convex composition}
\label{s3}

Yet another class of Jensen-like inequalities corresponds to
lower bounding the expectation of the
product of two functions, where one is convex and the other
is a composition of a positive monotonically non-decreasing function and a convex function, i.e.,
\begin{equation}
	\bE\{h[f(X)]g(X)\}\ge\sup_{a,b}\bE\{h[f(a)+f'(a)(X-a)]\cdot[g(b)+g'(b)(X-b)]\},
\end{equation}
where $f$ and $g$ are convex and $h$ is monotonically non-decreasing and non-negative.
For the case where $h(x)=e^x$, we end up with a bound that depends on the CGF of $X$ and its derivative:
\begin{eqnarray}
\bE\{e^{f(X)}g(X)\}&\ge&\bE\left\{e^{f(a)+f'(a)(X-a)}[g(b)+g'(b)(X-b)]\right\}\\
&=&e^{f(a)-af'(a)}\bE\left\{e^{Xf'(a)}[g(b)-bg'(b)+g'(b)X]\right\}\\
&=&\exp\{f(a)-af'(a)+\psi[f'(a)]\}\{g(b)+g'(b)(\psi'[f'(a)]-b)\}.
\end{eqnarray}
Maximizing w.r.t.\ $b$ while $a$ is kept fixed, yields $b_*=\psi'[f'(a)]$, and we obtain:
\begin{equation}
	\bE\{e^{f(X)}g(X)\}\ge \sup_a\exp\{f(a)-af'(a)+\psi[f'(a)]\}\cdot g(\psi'[f'(a)]).
\end{equation}

\noindent
{\em Example 5.}
Considering the case where $f(x)=-\ln x$ and $g(x)=x\ln x$, we may obtain a
reversed Jensen-like inequality, namely, a lower bound to the expectation of
the concave function $\ln X$:
\begin{eqnarray}
\bE\{\ln X\}&=&\bE\left\{e^{-\ln X}\cdot X\ln X\right\}\\
	&\ge&\sup_{a\ge 0}\exp\{-\ln a+1+\psi(-1/a)\}\cdot\psi'(-1/a)\ln\psi'(-1/a)\\
	&=&\sup_{\alpha\ge 0}\exp\{\ln\alpha+1+\psi(-\alpha)\}\psi'(-\alpha)\ln\psi'(-\alpha)\\
	&=&e\cdot\sup_{\alpha\ge 0} \alpha e^{\psi(-\alpha)}\psi'(-\alpha)\ln\psi'(-\alpha)\\
	&=&e\cdot\sup_{\alpha\ge 0} 
	\alpha \bE\{Xe^{-\alpha X}\}\ln\frac{\bE\{Xe^{-\alpha X}\}}{\bE\{e^{-\alpha X}\}}.
\end{eqnarray}
Defining the MGF $\phi(s)=\bE\{e^{sX}\}=e^{\psi(s)}$,
we have:
\begin{eqnarray}
	\bE\{\ln X\}&\ge&
	e\cdot\sup_{\alpha\ge 0}\alpha\phi'(-\alpha)\ln\psi'(-\alpha)\\
	&=&e\cdot\sup_{\alpha\ge 0}\alpha\phi(-\alpha)\psi'(-\alpha)\ln\psi'(-\alpha)\\
	&=&e\cdot\sup_{\alpha\ge 0}\alpha\phi'(-\alpha)\ln\frac{\phi'(-\alpha)}{\phi(-\alpha)}.
\end{eqnarray}
We obtained a lower bound
in terms of the MGF and its derivative (or, equivalently, the CGF and its derivative), which
is appealing in cases where $X$ is the sum of i.i.d.\ random variables.

\begin{figure}[h!t!b!]
\centering
\includegraphics[width=8.5cm, height=8.5cm]{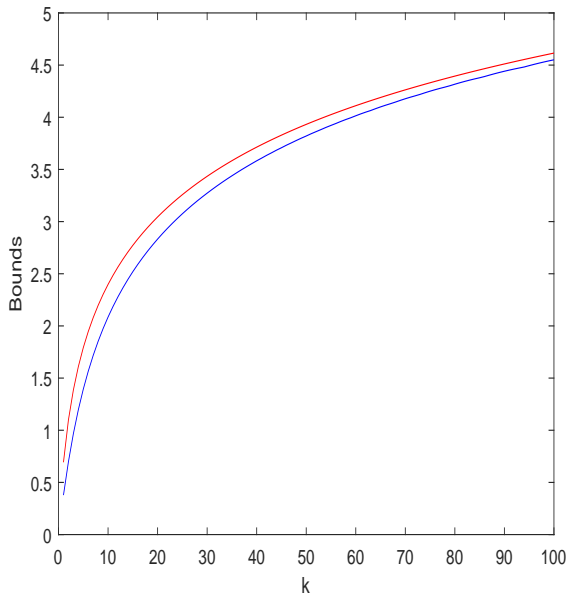}
\caption{Upper and lower bounds on
$\bE\left\{\ln\left(1+\sum_{i=1}^kY_i^2\right)\right\}$, where
$Y_i\sim\calN(0,\sigma^2)$ are i.i.d., for $\sigma^2=1$ and $k=1,2,\ldots,100$. The red curve is the
upper bound, $\ln(1+k\sigma^2)$, obtained by applying the ordinary 
Jensen inequality. The blue curve is the lower bound of
eq.\ (\ref{lb0}), where the search over $\alpha$ was carried out with
resolution of $0.001$.}
\label{graph1}
\end{figure}

Accordingly, we now particularize this example further by examining the case where
$X=1+\sum_{i=1}^kY_i^2$, with $Y_i\sim\calN(0,\sigma^2)$, $i=1,\ldots,k$,
being independent random variables. The
motivation of assessing an expression of the form,
$\bE\left\{\ln\left(1+\sum_{i=1}^kY_i^2\right)\right\}$, is two-fold. The
first is that it is
useful for bounding the ergodic capacity of the single-input,
multiple-output (SIMO) channel, where $\{Y_i\}$ designate random channel
transfer coefficients (see, e.g., \cite{me22}, \cite{DZWY15}, \cite{TV05} and
references therein). The second is that it is relevant for bounding
the joint differential entropy associated with the multivariate Cauchy
density. Here, $(Y_1,\ldots,Y_k)$ are not Gaussian as defined above,
but their multivariate Cauchy density can be represented as a continuous mixture of
i.i.d.\ zero-mean Gaussian random variables, where the mixture is taken over
all possible variances -- see \cite[Example 6]{me22} for the details.
In this case,
\begin{eqnarray}
	\phi(s)&=&\bE\left\{\exp\left(s\left[1+\sum_{i=1}^kY_i^2\right]\right)\right\}\\
	&=&e^s\left(\bE\{e^{sY^2}\}\right)^k\\
	&=&\frac{e^s}{(1-2s\sigma^2)^{k/2}},~~~~~~~s< \frac{1}{2\sigma^2}.
\end{eqnarray}
Thus,
\begin{equation}
	\psi(s)=s-\frac{k}{2}\ln(1-2s\sigma^2),
\end{equation}
and
\begin{equation}
	\psi'(s)=1+\frac{k\sigma^2}{1-2s\sigma^2}.
\end{equation}
It follows that
\begin{equation}
\label{lb0}
	\bE\left\{\ln\left(1+\sum_{i=1}^kY_i^2\right)\right\}\ge
	e\cdot\sup_{\alpha\ge 0}\left\{\frac{\alpha e^{-\alpha}}{(1+2\alpha\sigma^2)^{k/2}}
	\left(1+\frac{k\sigma^2}{1+2\alpha\sigma^2}\right)
	\ln\left(1+
	\frac{k\sigma^2}{1+2\alpha\sigma^2}\right)\right\}.
\end{equation}
The Jensen upper bound, $\ln(1+k\sigma^2)$,  and the lower bound (\ref{lb0}) are displayed in
Fig.\ \ref{graph1} for $\sigma^2=1$ and $k=1,2,\ldots,100$. As can be seen,
the bounds are quite close. Interestingly, the choice $\alpha=1/(k\sigma^2)$ yields 
results that are very close to those of the optimal $\alpha$.

\begin{figure}[h!t!b!]
\centering
\includegraphics[width=8.5cm, height=8.5cm]{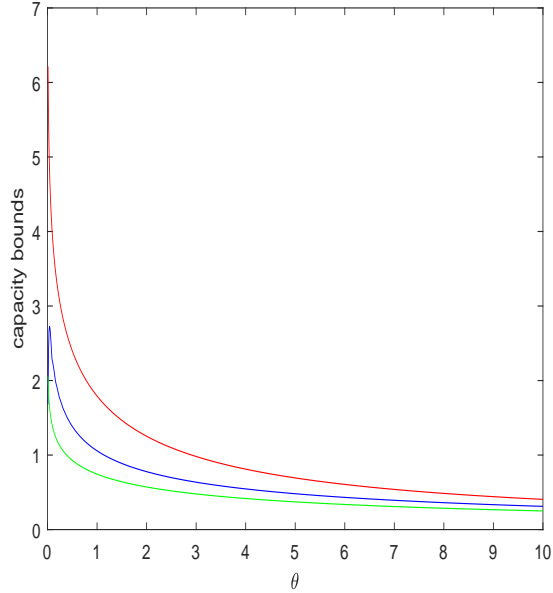}
\caption{Upper and lower bounds on
$\bE\left\{\ln(1+gZ)\right\}$, where $Z$ is distributed exponentially
with parameter $\theta$, as functions of $\theta$, for $g=5$.
The red curve is the
upper bound, $\ln(1+g/\theta)$, obtained by applying the ordinary
Jensen inequality. The blue curve is the lower bound of
of eq.\ (\ref{lb1}), where the search over $\alpha$ was carried out with
resolution of $0.001$. The green curve is the lower bound of
\cite[Example 1]{me22}.}
\label{graph2}
\end{figure}

Another instance of this example is the circularly symmetric complex
Gaussian channel
whose signal--to--noise ratio (SNR), $Z$, is a random variable (e.g., due to
fading), known to both the
transmitter and the receiver. The capacity is given by $C = \bE\{\ln(1 +
gZ)\}$, where $g$
is a certain
deterministic gain factor and the expectation is w.r.t.\
the randomness of $Z$. For
simplicity, let us assume that $Z$ is distributed exponentially, i.e.,
\begin{equation}
p(z)=\left\{\begin{array}{ll}
\theta e^{-\theta z} & z\ge 0\\
0 & z< 0\end{array}\right.
\end{equation}
where the parameter $\theta > 0$ is given. 
In this case,
\begin{equation}
\phi(-\alpha)=\frac{\theta e^{-\alpha}}{\theta+g\alpha}
\end{equation}
and
\begin{equation}
\psi(-\alpha)=\ln\theta-\ln(\theta+g\alpha)-\alpha,
\end{equation}
and so,
\begin{equation}
\label{lb1}
\bE\{\ln(1+gZ)\}\ge e\theta\cdot\sup_{\alpha\ge 0}\frac{\alpha
e^{-\alpha}}{\theta+g\alpha}\cdot\left(1+\frac{g}{g+\theta\alpha}\right)\ln\left(1+\frac{g}{g+\theta\alpha}\right).
\end{equation}
In Fig.\ \ref{graph2}, we plot this lower bound as a function of $\theta$ for
$g=5$ and compare it to the Jensen upper bound, $\ln(1+g/\theta)$ (red curve)
and to the lower bound of \cite[Sect.\ 4.1, Example 1]{me22}. As can be seen, the
lower bound proposed here is considerably tighter, especially for small
$\theta$.\\

\begin{figure}[h!t!b!]
\centering
\includegraphics[width=8.5cm, height=8.5cm]{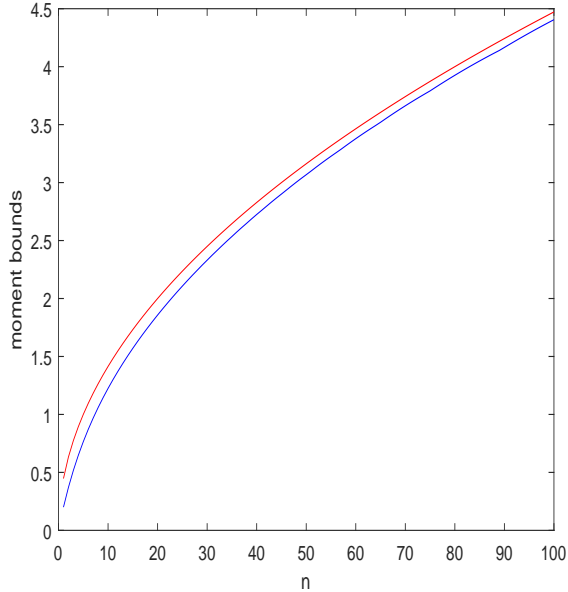}
\caption{Upper and lower bounds on
$\bE\{\sqrt{\sum_{t=1}^nY_t}\}$ as functions of $n$, where $\{Y_t\}$ are i.i.d., Bernoulli$(0.2)$.
The red curve is the Jensen upper bound, $\sqrt{np}$, and the blue curve is the
proposed lower bound where $\alpha$ is optimized in the range $[0,10]$ and $s$
is optimized in the range $[0.5,10]$, both with resolution of $0.01$.}
\label{graph3}
\end{figure}

\noindent
{\em Example 6.}
Yet another example of this family of Jensen-like inequalities, applies to obtaining a lower bound to
$\bE\{X^t\}$, where $t$ is an {\em arbitrary} real. For a given $t$, let $s\ge
0$ be either larger than $1-t$ or smaller than $-t$, consider the case where
$f(x)=x^{t+s}$, $g(x)=-s\ln x$ and $h(x)=e^x$. Then,
\begin{eqnarray}
\bE\{X^t\}&=&\bE\{e^{-s\ln X}X^{t+s}\}\\
&\ge&\bE\left\{\exp\left[s\left(-\ln
a-\frac{1}{a}(X-a)\right)\right]\cdot\left[b^{t+s}+(t+s)b^{t+s-1}(X-b)\right]\right\}\\
&=&e^{s[1-\ln
a]}\phi\left(-\frac{s}{a}\right)\left[b^{t+s}+(t+s)b^{t+s-1}\left(\psi'\left(-\frac{s}{a}\right)-b\right)\right].
\end{eqnarray}
Choosing $b=\psi'(-s/a)$, and changing the optimization variable $a$ into
$\alpha=1/a$, we get
\begin{equation}
\bE\{X^t\}\ge \sup_{\alpha\ge 0}(\alpha e)^s\phi(-\alpha
s)[\psi'(-\alpha s)]^{t+s}.
\end{equation}
More specifically, if $X=\sum_{i=1}^nY_i$, where $\{Y_i\}$ are Bernoulli i.i.d.,
with parameter $p$, then $\phi(s)=(pe^s+q)^n$, where $q=1-p$. We then obtain
\begin{equation}
\bE\{X^t\}\ge \sup_{\alpha\ge 0}(\alpha e)^s(pe^{-\alpha
s}+q)^n\cdot\left(\frac{npe^{-\alpha s}}{pe^{-\alpha s}+q}\right)^{t+s}.
\end{equation}
Selecting $\alpha=1/(np)$, we obtain
\begin{equation}
\bE\{X^t\}\ge
(np)^t\cdot\frac{e^s(pe^{-s/(np)}+q)^ne^{-s(t+s)/(np)}}{(pe^{-s/(np)}+q)^{t+s}}.
\end{equation}
The first factor is $(\bE X)^t$. The second factor tends to unity as $n$ grows,
because $pe^{-s/np}+q\approx p(1-s/(np))+q=1-s/n$, and so,
$(pe^{-s/np}+q)^n\approx (1-s/n)^n\approx e^{-s}$. For $t\ge 1$ and $t\le 0$,
the function $f(x)=x^t$ is convex, and so, $(\bE X)^t$ is the ordinary Jensen lower bound.
In this case, the bound is valuable if the multiplicative factor, 
$$\frac{e^s(pe^{-s/(np)}+q)^ne^{-s(t+s)/(np)}}{(pe^{-s/(np)}+q)^{t+s}},$$
is larger than unity. If $0<t<1$, the function $f(x)=x^t$ is concave, and then $(\bE
X)^t$ is an upper bound. Of course, the parameter $s$ can be optimized too.
Some numerical results for $t=0.5$ are depicted in Fig.\ \ref{graph3}.
As can be seen, the upper and the lower bounds are fairly close.

\begin{figure}[h!t!b!]
\centering
\includegraphics[width=8.5cm, height=8.5cm]{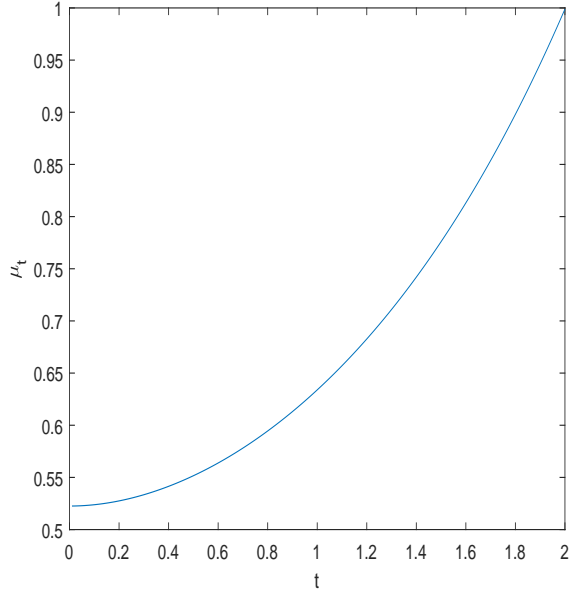}
\caption{The gap factor, $\mu_t$, as a function of $t$.
The parameter $s$
is optimized in the range $[1-t/2,10]$ with resolution of $0.001$.}
\label{graph4}
\end{figure}

Another application of this example is related to estimation theory.
Let $\theta\in\reals$ and let $Y_1,\ldots,Y_n$ be i.i.d.,
with mean $\theta$ and variance $\sigma^2$. Consider the 
$t$-th moment of the estimation error,
$\bE_\theta\bigg|\frac{1}{n}\sum_{i=1}^nY_i-\theta|^t$.
Defining $X=\left(\frac{1}{n}\sum_{i=1}^nY_i-\theta\right)^2$,
we have
\begin{equation}
\phi(s)=\frac{1}{\sqrt{1-2s\sigma^2/n}};~~~
\psi(s)=-\frac{1}{2}\ln\left(1-\frac{2s\sigma^2}{n}\right).
\end{equation}
and so,
\begin{equation}
\phi(-\alpha s)=\frac{1}{\sqrt{1+2\alpha s\sigma^2/n}};~~~~
\psi'(-\alpha s)=\frac{\sigma^2/n}{1+2\alpha s\sigma^2/n}.
\end{equation}
\begin{eqnarray}
\bE_\theta\bigg|\frac{1}{n}\sum_{i=1}^nY_i-\theta\bigg|^t&=&\bE_\theta X^{t/2}\\
&\ge&\frac{(\alpha e)^s}{\sqrt{1+2\alpha s\sigma^2/n}}\left(\frac{
\sigma^2/n}{1+2\alpha s\sigma^2/n}\right)^{t/2+s}\nonumber\\
&=&\left(\frac{\sigma^2}{n}\right)^{t/2+s}\cdot\frac{(\alpha e)^s}{(1+2\alpha
s\sigma^2/n)^{(t+1)/2+s}}.
\end{eqnarray}
with either $s\ge 1-t/2$ or $s\le-t/2$.
For $\alpha=\zeta n/\sigma^2$ ($\zeta> 0$ being a constant), we have:
\begin{equation}
\bE_\theta\bigg|\frac{1}{n}\sum_{i=1}^nY_i-\theta\bigg|^t\ge
\frac{\sigma^t}{n^{t/2}}\cdot\sup_{\zeta>0,~s>1-t/2}\frac{(\zeta e)^s}{(1+2\zeta s)^{(t+1)/2+s}}
\end{equation}
where for $t\in[0,2]$, the first factor, $\sigma^t/n^{t/2}$, is the Jensen
upper bound. The second factor, 
\begin{equation}
\mu_t=\sup_{\zeta>0,~s>1-t/2}\frac{(\zeta e)^s}{(1+2\zeta s)^{(t+1)/2+s}},
\end{equation}
is the gap between the Jensen upper bound and the proposed lower
bound. In Fig.\ \ref{graph4}, we display this factor. The result $\mu_2=1$
is expected, because for $t=2$ and $s=0$, the calculation of is trivially
exact. Note that the maximization over $\zeta$, for a given $s$,
can be carried out in closed form, by equating to zero the partial derivative
of $\ln[(\zeta e)^s/(1+2\zeta s)^{(t+1)/2+s}]$ w.r.t.\ $\zeta$. The optimal
$\zeta$ turns out to be equal to $1/(t+1)$ (independently of $s$), and so,
\begin{equation}
\mu_t=\sup_{s>1-t/2}\left(\frac{t+1}{t+2s+1}\right)^{(t+1)/2}\cdot\left(\frac{e}{t+2s+1}\right)^s.
\end{equation}

Finally, it should be pointed out that this family of Jensen-like bounds,
opens the door also to
lower bound calculations on the form $\bE\{f(X)/g(X)\}$, where $f$ is positive
convex and $g$ is positive and concave. Using the fact the identity,
$1/s=\int_0^\infty e^{-st}\mbox{d}t$, we have:
\begin{eqnarray}
\bE\left\{\frac{f(X)}{g(X)}\right\}&=&\bE\left\{f(X)\cdot\int_0^\infty
e^{-tg(X)}\mbox{d}t\right\}\\
&=&\int_0^\infty\bE\left\{e^{-tg(X)}f(X)\right\}\mbox{d}t
\end{eqnarray}
and we can apply the same ideas as before to the integrand, having the freedom to
optimize the bound parameters with possible dependence on $t$.

\section{A product of two positive convex functions}
\label{s4}

The last family of Jensen--like bounds that we present in this work is
associated with the product of two non-negative convex functions.
Let both $f$ and $g$ be non-negative convex functions of $x\ge 0$. Then,
\begin{eqnarray}
\bE\{f(X)g(X)\}&\ge&\bE\{[f(a)+f'(a)(X-a)]\cdot g(X)\}\\
&=&[f(a)-af'(a)]\bE\{g(X)\}+f'(a)\bE\{Xg(X))\}\\
&\ge&[f(a)-af'(a)]\bE\{[g(b)+g'(b)(X-b)]\}+\nonumber\\
& &f'(a)\bE\{X[g(c)+g'(c)(X-c)]\}~~~~~~~f(a)\ge
af'(a)\ge 0\\
&=&[f(a)-af'(a)]\cdot[g(b)-bg'(b)+g'(b)\bE\{X\}]+\nonumber\\
& &f'(a)[(g(c)-cg'(c))\bE\{X\}+g'(c)\bE\{X^2\}\}].
\end{eqnarray}
The optimal $b$ and $c$ are $b^*=\bE\{X\}$ and
$c^*=\bE\{X^2\}/\bE\{X\}$, respectively. Thus,
\begin{equation}
\bE\{f(X)g(X)\}\ge[f(a)-af'(a)]\cdot g(\bE\{X\})+f'(a)\bE\{X\}\cdot
g\left(\frac{\bE\{X^2\}}{\bE\{X\}}\right).
\end{equation}
Let  
\begin{equation}
a^*=\frac{\bE\{X\}\cdot g(\bE\{X^2\}/\bE\{X\})}{g(\bE\{X\})}
\end{equation}
and assume that $f(a^*)\ge a^*f'(a^*)\ge 0$. Then $a^*$ is the optimal value
of $a$, which yields
\begin{equation}
\bE\{f(X)g(X)\}\ge f\left(\frac{\bE\{X\}\cdot
g(\bE\{X^2\}/\bE\{X\})}{g(\bE\{X\})}\right)\cdot g(\bE\{X\}).
\end{equation}
More generally, when $X$ and $Y$ are two random variables with a joint
distribution, the above derivation easily extends to
\begin{equation}
\bE\{f(X)g(Y)\}\ge f\left(\frac{\bE\{X\}\cdot
g(\bE\{XY\}/\bE\{X\})}{g(\bE\{Y\})}\right)\cdot g(\bE\{Y\}).
\end{equation}
If $f$ and $g$ are both concave, rather than convex, then the inequalities are
reversed.\\

\noindent
{\em Example 7.}
Consider again the example of the capacity of the AWGN with a random SNR,
$c(Z)=\ln(1+gZ)$, and suppose that we wish to bound the variance of $c(Z)$
in order to assess the fluctuations (e.g., for the purpose of bounding the
outage probability).
Then, obviously,
\begin{equation}
\mbox{Var}\{c(Z)\}=\bE\{c^2(Z)\}-[\bE\{c(Z)\}]^2=\bE\{\ln^2(1+gZ)\}-[\bE\{\ln(1+gZ)\}]^2.
\end{equation}
To upper bound $\mbox{Var}\{c(Z)\}$, we may derive an upper bound to
$\bE\{\ln^2(1+gZ)\}$ and a lower bound to $\bE\{\ln(1+gZ)\}$. For the latter, a
lower bound was already proposed earlier in Example 5. For the former, we may use the
present inequality with the choice $f(z)=g(z)=\ln(1+gz)$, which can easily be
shown to satisfy the requirements. We then obtain
the following upper bound, which
depends merely on the first two moments of $Z$:
\begin{equation}
\bE\{\ln^2(1+gZ)\}\le
\ln(1+g\bE\{Z\})\cdot\ln\left(1+\frac{g\bE\{Z\}\ln(1+g\bE\{Z^2\}/\bE\{Z\})}{\ln(1+g\bE\{Z\})}\right).
\end{equation}
Interestingly, the function $\ln^2(1+gx)$ is neither convex nor concave, yet
our approach offers an upper bound, which is fairly easy to calculate provided
that one can compute the first two moments of $Z$.

\end{document}